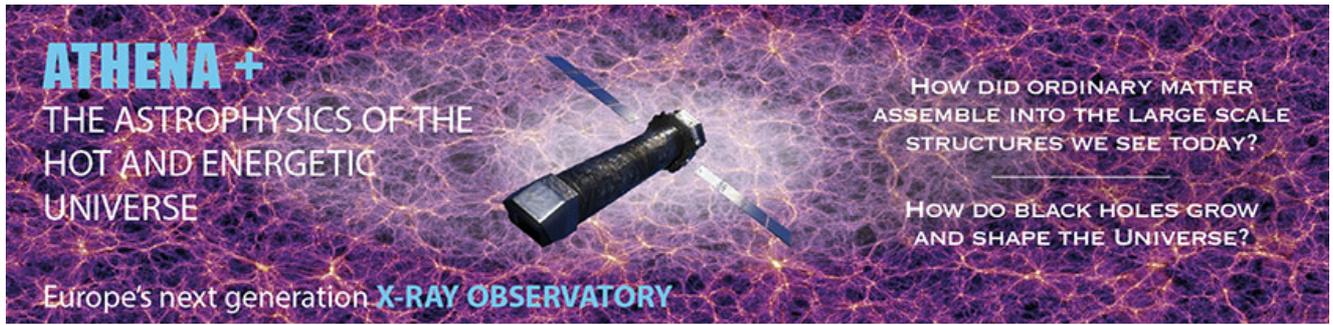

# The Hot and Energetic Universe

An *Athena+* supporting paper

# The missing baryons and the Warm-Hot Intergalactic Medium

## Authors and contributors

**Jelle Kaastra, Alexis Finoguenov,** Fabrizio Nicastro, Enzo Branchini, Joop Schaye, Nico Cappelluti, Jukka Nevalainen, Xavier Barcons, Joel Bregman, Judith Croston, Klaus Dolag, Stefano Ettori, Massimiliano Galeazzi, Takaya Ohashi, Luigi Piro, Etienne Pointecouteau, Gabriel Pratt, Thomas Reiprich, Mauro Roncarelli, Jeremy Sanders, Yoh Takei, Eugenio Ursino

The Hot and Energetic Universe: The missing baryons and the Warm-Hot Intergalactic Medium

# 1. EXECUTIVE SUMMARY

The backbone of the large-scale structure of the Universe is determined by processes on a cosmological scale and by the gravitational interaction of the dominant dark matter. However, the mobile baryon population shapes the appearance of these structures. Details about how this happens are often poorly known. Theory predicts that most of the baryons reside in vast unvirialized filamentary structures that connect galaxy groups and clusters (the "Cosmic Web"), and this is confirmed by measurements of the baryon density in the Ly$\alpha$ forest at $z>2$. But at the current epoch ($z<1$) about one half of the baryons are missing. Because the majority of the baryons are supposed to exist in a large-scale, hot and dilute gaseous phase, X-rays are the ideal tool to search for and study them. Sensitive, high-resolution X-ray observations, such as the ones that will be provided by *Athena+,* will teach us about the formation, structure and evolution of these largest objects of the Universe.

At present, our best observational constraints stem from the cooler gas accessible to the optical and UV band or from the densest structures represented by clusters of galaxies and the heavier groups of galaxies. The warm-hot intergalactic medium (WHIM), the haloes of galaxies and the cluster outskirts connected to the cosmic web are not so well constrained by observations and hence there are many unanswered questions.

Key questions that *Athena+* will address are:

- Where are the baryons still missing from the cosmic budget at $z<1$? Do they really trace the filaments of the cosmic web, as the theory predicts? What is their physical state and composition?
- Where have the missing baryons in galactic haloes, including our own Galaxy, gone? Do we see them in the circumgalactic space?
- What is the role of feedback by galactic winds and active galactic nuclei in the process of galaxy formation?
- What is the fate of the gas? How much material is accreted, how much is blown out, and what fraction is locked-up temporarily in stars?
- What are the relative contributions from accretion versus outflows in structure formation?

Such questions are key to understand what is the role of feedback by star-driven galactic winds and active galactic nuclei in the process of galaxy formation and what is the balance between gas accretion and outflows (see Cappi, Done et al., 2013, *Athena+* support paper and Georgakakis, Carrera et al., 2013, *Athena+* support paper).

**Observations with *Athena+* will reveal the location, chemical composition, physical state and dynamics of the active population of baryons.**

# 2. INTRODUCTION

Over the past 15 years our understanding of the Universe as a whole has progressed toward what we now define as a precision cosmology that establishes, with an accuracy better than half a percent, that we live in a flat and mysterious Universe, where the ordinary baryonic matter makes up only 5% of its mass-energy budget and the remaining 95% is dubbed 'dark' mainly to reflect our inability to directly detect and identify it (e.g. Komatsu et al. 2009; Planck collaboration 2013). However, if the general picture is now reasonably understood and proven to a high degree of accuracy, the details remain largely unknown: very little is known about the nature and origin of 'dark' Energy (70%) and 'dark' Matter (25%), and the situation is only conceptually better for the remaining 5% of baryons, about half of which is still elusive at $z=0$.

**Baryons are missing on cosmological scales:** in the early Universe (<2 billion years from the 'Big Bang') most of the baryons were relatively cold and homogeneously distributed, and therefore easily detectable through redshifted HI absorption in the optical spectra of background quasars ('Ly$\alpha$ Forest' studies, e.g. Rauch 1998, Weinberg 1997). As time progressed, however, density perturbations in the Universe grew nonlinearly and their physics became more complex. Structures began forming rapidly and baryons were accreted and expelled from galaxies, heating-up through shocks while collapsing to form structures, cooling down to form stars in galaxies, getting enriched with metals while shining in stars, and finally being recycled out of galaxies via powerful supernova- and AGN-driven outflows. As a result of these continuing violent processes of structure formation, **today's observations can only account for <60% of the baryons that our precision cosmology predicts in our local Universe, and that have been detected in the high-z, linear Universe** (e.g. Fukugita 2003; Shull et al. 2012).

**But baryons are missing not only at the largest scales: most galaxies fall short of baryons when their measurable baryonic fraction is compared with the universal baryon to total mass ratio** (e.g. Bell et al. 2003; McGaugh 2010). The problem is more severe for smaller galaxies, which suggests that shallower potential well haloes fail to retain baryons during mergers and/or bursts of star formation activity. If so, the baryons missing from galaxies should be found in the circum-galactic space (CGM), probably even beyond their virial radii.

The two missing-baryon problems are obviously related and theoretically reconciled by the leading Λ-CDM theory of cosmological structure formation. This theory includes feedback processes that predict that these 'missing' baryons are hiding in a tenuous warm-hot intergalactic matter. This gas, the so-called Warm-Hot Intergalactic Medium (WHIM), is shock-heated to temperatures of $10^5$-$10^7$ K and is structured in the form of a web whose 'nodes' host the gaseous outskirts (i.e. circum-galactic medium and circum-cluster medium) of the Universe's virialized structures (galaxies and galaxy clusters), e.g. Cen & Ostriker (2006). These baryons are too hot to be visible at energies lower than X-rays, and too tenuous to be detectable with current X-ray instrumentation.

Studying the physics, chemistry and dynamics of the Universe's hot and roaming baryons, as a function of the cosmic time and distance from virialized structures, will give us a unique set of tools to (1) study the evolution of Large Scale Structures in the Universe, (2) map the outskirts of dark matter concentrations and (3) identify the different feedback mechanisms at work, and their role in affecting the growth of structures, inhibiting star formation, metal enrichment, heating, and eventually re-cooling to become the inter-galactic medium (IGM).

## 3. FILAMENTS AND LARGE-SCALE STRUCTURES

### 3.1. Baryons in the cosmic web

Theory predicts that a large fraction of the baryons preferentially populate the large filamentary structures that constitute the backbone of the "Cosmic Web". Evidence for these baryons has been obtained via UV-absorption line studies with FUSE and HST-COS (Shull et al. 2012). These observations probe the coldest fraction of these baryons while, according to hydrodynamical simulations (e.g. Cen & Ostriker 2006, Branchini et al. 2009, Bertone et al. 2010, Smith et al. 2011, Roncarelli et al. 2012), the majority lie in a hotter phase (i.e. T>$10^{5.5}$ K) that can only be detected through highly ionized C, N, O, Ne, and Fe ions in X-rays.

While reconciling the baryon budget in the local universe with that in the earlier epochs is still one of the main goals of cosmology, the wealth of astrophysical information encoded in the WHIM makes its study of the utmost importance.

Baryons in filaments are shock-heated during the buildup of large-scale structures. An additional heating source, which becomes progressively more important towards higher density regions, is provided by the stellar feedback that also determines the metal abundance in the WHIM. Thanks to their remoteness from external influences and their relative simple physics, these baryons represent a unique probe of structure formation processes and metal enrichment history, and their study can shed light on the complex interplay between galaxies and the IGM, through both accretion and feedback.

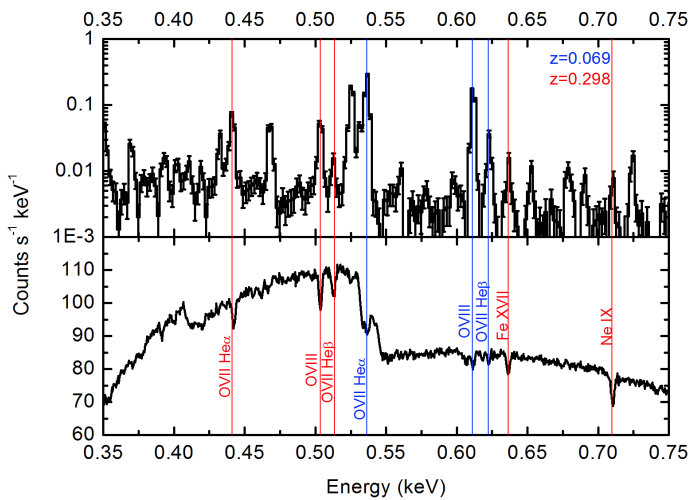

Figure 1: Simulated emission and absorption line spectra captured in a single *Athena+* observation for two filaments at different redshifts. Lower panel: absorption spectrum from a sight line where two different filamentary systems are illuminated by a bright background source. Upper panel: corresponding emission from a 2'x2' region from the same filaments for 1 Ms exposure time. The high spectral resolution allows us to distinguish both components. *Athena+* will be able to study dozens of these sight lines in detail.

Thanks to its unique combination of energy resolution, line sensitivity and imaging capabilities, the Athena+ X-IFU can achieve all these goals by measuring the WHIM lines both in absorption against a bright distant source and in emission (Fig. 1).

## 3.2. Absorption line studies

Current results do not yet provide a convincing evidence for these baryons in X-ray absorption spectra (Nicastro et al. 2013 and reference therein). Therefore our first, obvious goal is their unambiguous detection. Thanks to the large number of expected detections along many independent lines of sight, which minimize the impact of cosmic variance, these measurements can set strong constraints on the WHIM properties, whereas a non-detection would seriously undermine our scenario of baryonic structure formation.

We envision a programme consisting of observations of about 25 AGN and 40 GRB afterglows, totaling about 7 Ms, and yielding about 200 filaments in 5 years. Pushing the resolution to 1 eV would improve the limiting sensitivity, dominated by systematics, more than doubling the detection rate with a 50% increase in the total time.

The power of this programme is illustrated in Fig. 2 in which the expected O VII line counts as a function of equivalent width (EW) accumulated in 5 years (black dots) are superimposed on theoretical predictions (colored band). The width of the band indicates the current spread in theoretical predictions, while different colors highlight the EW regions accessible to *Athena+* assuming different energy resolutions of the X-IFU. Thanks to the small errors driven by Poisson noise, simple line counts will be able to discriminate efficiently among different models, including the pre-enrichment scenarios advocated to justify the large metal abundance the cluster outskirts (Matsushita et al. 2013) that occupy the upper part of the band in the plots. Residual degeneracy among models can be removed by considering line counts in different redshift bins, hence it will be important to distant GRBs and faint AGNs in addition to the bright, local ones.

To determine the cosmic abundance of these baryons one needs to estimate the absolute metallicity of the medium, which can be obtained from the X-ray spectra of UV-bright AGN where broad Lyman-α absorbers have been previously measured by HST-COS. Once again, a large number of possible targets and detections accessible to *Athena+* will be crucial to minimize cosmic variance and obtain a reliable estimate of the mean cosmic density of these baryons.

O VII and O VIII are the best tracers of baryons in filaments. In 30-50% of cases *Athena+* will also detect additional metal lines whose importance in determining the physical state of the baryons has been outlined by the recent analysis of the *Chandra* X-ray spectrum of 1ES 1553+113 (Nicastro et al. 2013).

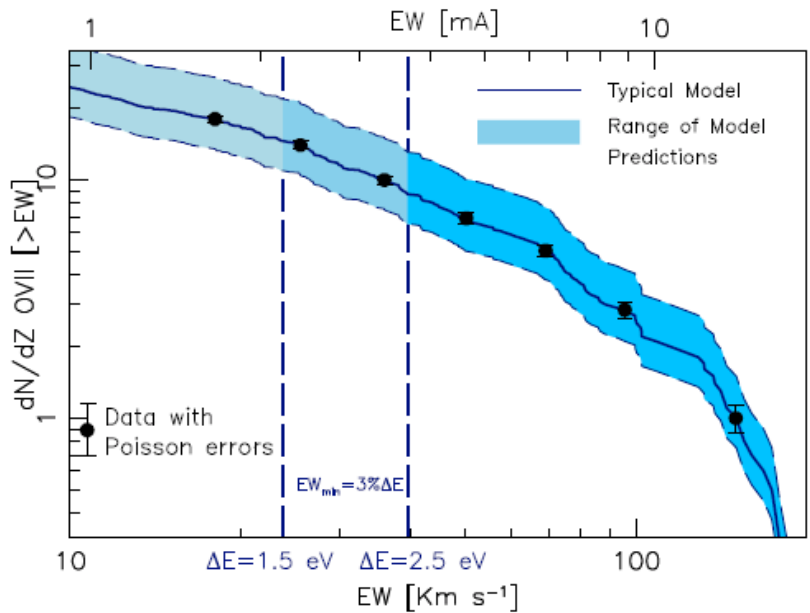

Figure 2: Colored band: the predicted mean number of O VII absorption lines per unit redshift as a function of line EW (Cen & Fang 2006, Branchini et al. 2009). Different colors highlight the EW ranges above the minimum value that can be detected by Athena+. The minimum EW is set by systematic effects and depends on the energy resolution, as indicated in the plot, and assumes 1% uncertainty on the spectral continuum

## 3.3. Emission line studies

Emission lines are significantly weaker and more difficult to detect. However, *Athena+* is expected to detect emission lines for a sizable (~30%) fraction of the systems identified in absorption. Simultaneous absorption and emission spectroscopy will allow for direct, model-independent measurement of the gas density, length scale, ionization balance, excitation mechanism (or gas temperature), and element abundance. Emission lines preferentially probe the gas in high-density environments, bridging the gap to the clusters of galaxies and potentially tracing the gas infall onto virialized objects. Fig. 3 shows the same statistics as Fig. 2 but for emission lines as a function of the line surface brightness. Data points show simulated line count expectations after 5 years, corresponding to several hundreds of filaments mostly from serendipitous detections in the field of view of high Galactic latitude sources. Error bars account for (dominant) shot noise errors and assume negligible cosmic variance. The upper band illustrates the typical spread in theoretical predictions (Cen & Fang 2006, Takei et al. 2011) and does not account for more conservative models (see e.g. Roncarelli et al. 2012) since these are probably in tension with metal abundance in the outer part of galaxy clusters. Baryons in filaments (the lower band) constitute a fraction of the detected lines that, as anticipated, will preferentially probe regions of higher density. Efficient separation of the two phases will be possible thanks to X-IFU angular resolution that will allow to remove the contribution from compact and point-like sources.

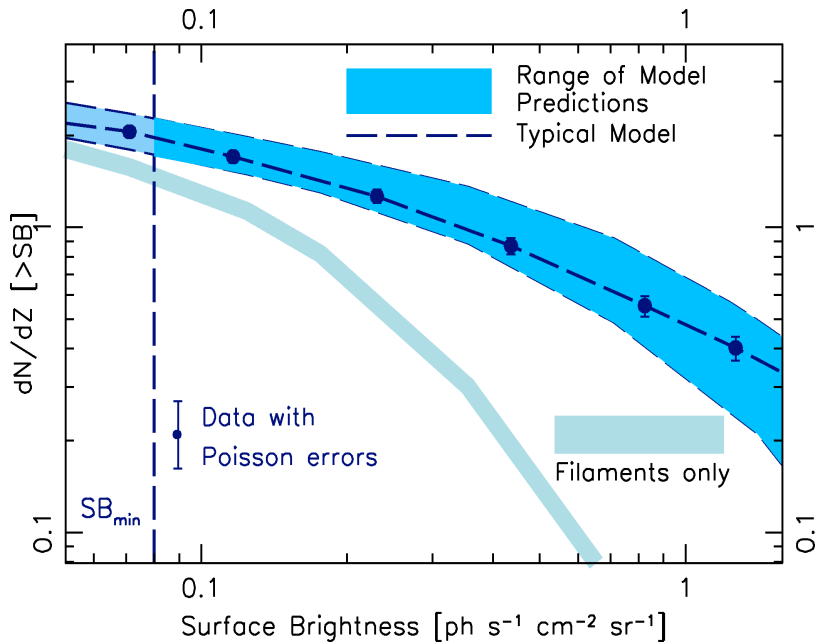

Figure 3: Same line statistics as in Fig. 2 but referring to filaments detected in emission as a function of line surface brightness (Takei et al. 2011). We expect to detect serendipitously about 4 filaments in the field of view of X-IFU for observations longer than 100 ks. We have conservatively assumed that for about 30% of the systems detected in absorption, the associated X-ray line emission will be detected. The upper band illustrates the scatter in model predictions and refers to simultaneous detections of O VII and O VIII from all the baryons. The lower band refers to line emission from baryons in filaments only. This component increasingly dominates low SB. For observations longer that 100 ks, we expect that about 50% of the filaments detected are due to the WHIM.

These predictions, both in emission and absorption, assume simultaneous detection of at least two lines, e.g. O VII and O VIII, at the same redshift. The width of the band in Fig. 3 indicates the spread in current theoretical predictions (Takei et al. 2011) with highly localised metal diffusion models occupying the lower part of the stripe and diffusion into the IGM via SN feedback-related processes populating the upper part. The programme related to emission studies is based on serendipitous observations. Assuming that about 30% of the observing programme will go to long X-IFU observations (>100 ks), we predict about thousand detections. Similar to the absorption case, pushing the resolution to 1 eV improves the limiting sensitivity, dominated by systematics, more than doubling the detection rate with a 50% increase in the total time.

## 3.4. Gamma Ray Bursts behind walls

All absorption line studies discussed before are based on bright background sources available by chance (AGN, GRBs) and not biased towards any particular line of sight. There can be fortunate circumstances, however, that a bright background source is found behind one of the most pronounced cosmic web structures. At present, one such case is known, the blazar H2356-309 behind the Sculptor Great Wall, a large-scale superstructure of galaxies at z~0.03. The combination of 600 ks *Chandra-LETGS* and 130 ks *XMM-Newton-RGS* grating spectra showed an O VII absorption line from the wall at the 4σ significance level (Fang et al. 2010).

Observing GRB afterglows when they occur behind such structures offers a unique opportunity to complement absorption diagnostics with emission studies of bridges between clusters. Fig. 4 shows that it is well possible to detect absorption by a Sculptor-like structure of a GRB X-ray afterglow, when it is observed with *Athena+* for 50 ks, starting 2 hours after the initial burst. In that case a 3σ detection of the O VII absorption line is obtained for oxygen columns of $10^{15}$ cm$^2$, i.e. that expected for typical WHIM filaments.

If the observation starts 8h after the initial burst, 3σ detections are still possible, but require higher oxygen column densities ($10^{15}$-$10^{16}$ cm$^{-2}$) and thus a smaller fraction of WHIM filaments can be probed and the detection probability becomes smaller.

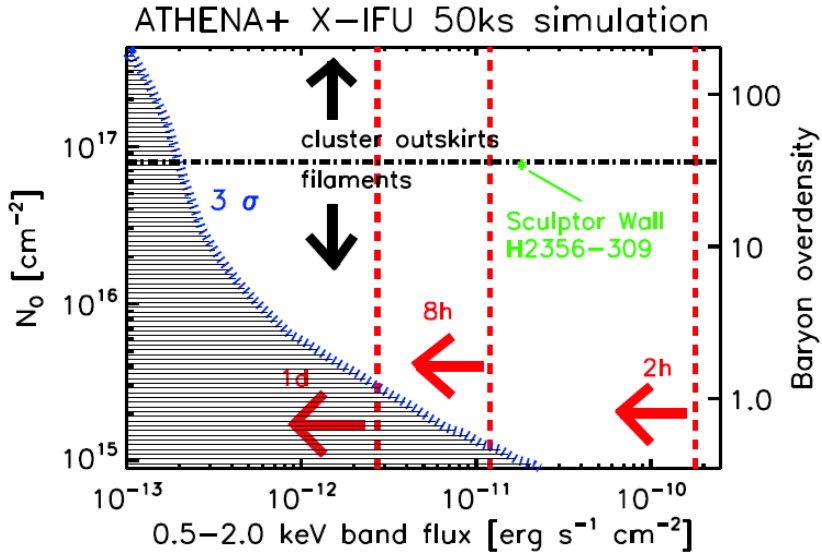

Figure 4: The blue dotted line shows the 3σ detection limit for *Athena+* for OVII absorption lines for a range of oxygen column densities $N_O$ and average background flux values (the shaded area shows the combination of low fluxes and $N_O$ values which yield too low detection). Exposure time 50 ks, WHIM temperatures and oxygen abundances as for the Sculptor event (Fang et al. 2010). Red lines and arrows indicate the maximum values of the Gamma Ray Burst X-ray afterglow fluxes as measured with Swift (Evans et al. 2009), as a function of elapsed time after the initial burst. The dash-dot line divides roughly the over-density space between the galaxy cluster outskirts and the large-scale filaments.

### 3.5. Cluster outskirts

The cosmic web filaments discussed before are connected to clusters of galaxies, and reach their highest intensity near the cluster outskirts. **Cluster outskirts are a melting pot, where cosmic baryons convert their kinetic energy of the infall into the thermal energy of the hot gas.** Studies of these zones in emission and absorption have just become feasible with current instrumentation, indicating that *Athena+* will be able to provide a complete picture there. The state of the baryons is expected to show a wide spectrum of thermal and kinematic properties. **This is where the high-energy resolution, high collecting area and large field of view studies are required, making *Athena+* an unprecedented instrument that will reveal the full picture of baryons on large scales.**

*Athena+* will have the capabilities to study these zones both in emission and in absorption, which will allow a full 3D picture of the gas properties to be created. *Athena+* studies will reveal the classical effects of the infall of baryons, such as the Kaiser effect (distortions from the Hubble flow caused by the velocity fields of the infalling matter, see also Fig. 5), that has so far been traced only by collisionless tracers, the galaxies (Ursino et al. 2011). Using gas instead of galaxies as proxy will enable direct investigation of the dynamical and thermal state of collisional baryons at larger distances from clusters.

Studies with high resolution are a unique way to peer into faint emission, revealing only the parts of the cosmic web at the redshift of interest and effectively avoiding fore- and backgrounds. These studies will be accompanied by a detailed reconstruction of the dark matter content of the cosmic web by *Euclid* (Laureijs et al. 2011).

### 3.6. Correlation studies

The unresolved Cosmic X-ray Background contains information about all those sources that have not been detected at the deepest fluxes reachable by deep surveys. The amplitude of its power spectrum is not only sensitive to the luminosity density of those sources, but it also provides information about their bias.

Although it has very low surface brightness, the WHIM can be distinguished from other kinds of diffuse emission on the basis of its clustering properties (Ursino & Galeazzi 2006). In fact, its emission follows that of clusters and filaments and peaks at low redshift (z~0.5), thus showing a typical feature in the angular clustering of unresolved CXB fluctuations. However, the WHIM is not the only expected component of the unresolved CXB arising from thermal emission of the Inter Galactic Medium (IGM). X-ray surveys in the local Universe revealed X-ray emission from local galaxy groups down to masses of the order of $10^{12}$ M$_\odot$ (see e.g. Eckmiller et al. 2011). Since the intensity of the X-ray emission of galaxy groups scales with their mass, a large fraction of them has not been detected at moderately high-z. Therefore we also expect a contribution to the overall signal from medium to low mass-groups at z>0.2-0.3.

Concerning the unresolved extragalactic CXB, it is difficult to distinguish its components with a simple spectral analysis mostly because of the poor energy resolution of X-ray sensitive CCDs. However, cosmological sources leave a unique imprint in the power spectrum of the anisotropies of the fluctuations of the unresolved CXB in a way that is related to their clustering and volume emissivity properties.

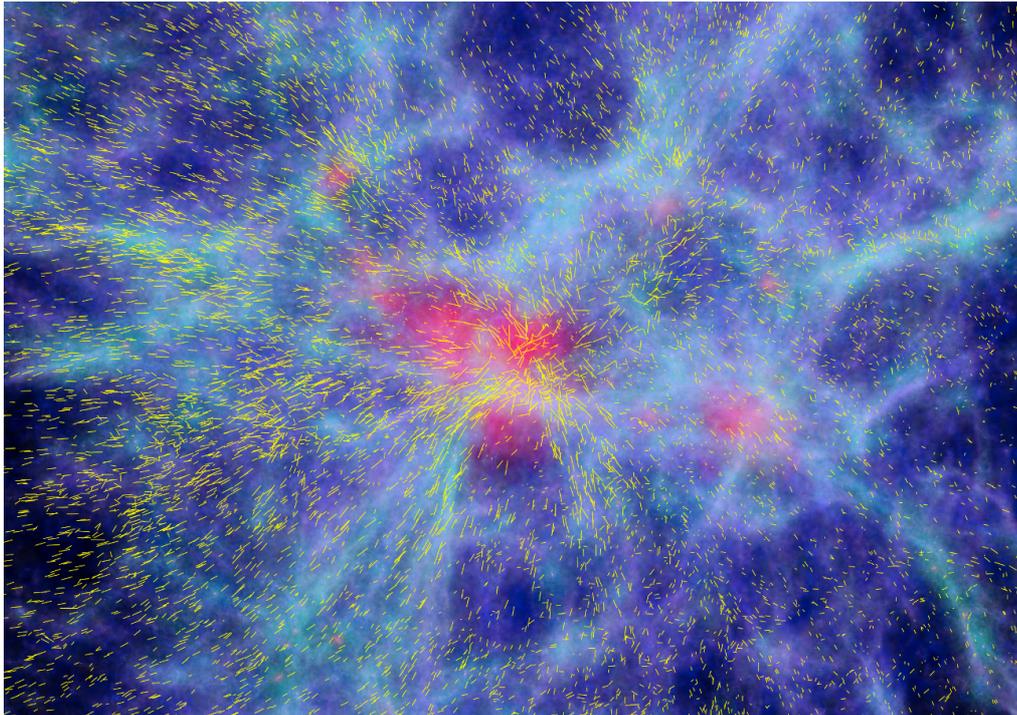

Figure 5: Snapshot from a computer simulation of the formation of large-scale structures in the Universe, showing a patch of 30 Mpc across and the resulting coherent motions of galaxies flowing towards the highest mass concentration in the centre. The snapshot refers to an epoch ~$10^{10}$ years back in time. The colour scale represents the mass density (highest density regions red, lowest density black). The tiny yellow lines show the intensity and direction of the galaxy's velocities. Like compass needles, they map the infall pattern and measure the rate of growth of the central structure. This depends on the subtle balance between dark matter, dark energy and the expansion of the Universe. Courtesy of Klaus Dolag

Using several *XMM-Newton* observations, Galeazzi et al. (2009) were able to identify the WHIM autocorrelation signal in the 0.4-0.6 keV band. Although they could not set any constraint on models, they could estimate the WHIM contribution to 12±5% of the DXB in this band. In a recent paper Cappelluti et al. (2012) showed that at the flux limit of the *Chandra* deep fields the main contributor to the power spectrum of the unresolved CXB is the emission from diffuse Warm-Hot sources. However because of the limited field of view of *Chandra*, such a measurement was only qualitative and did not provide any constraints for the models of WHIM and low surface brightness emission sources and their clustering. *Athena+* will be the step forward for such a study; its field of view allows the CXB fluctuation to be studied out to scales of 0.5° in a single pointing.

Moreover, given its sensitivity to low surface brightness emission and point sources, it will be possible to remove foreground sources in an unprecedented way in order the study the clustering properties of WHIM by avoiding contamination from other sources producing the unresolved CXB.

The Hot and Energetic Universe: The missing baryons and the Warm-Hot Intergalactic Medium# 4. GALACTIC HALOES

Although many of the ingredients of the theory of the formation and evolution of galaxies are in place, the process is far from understood. Chief among the most important, poorly understood processes are gas accretion and galactic winds.

Models predict that the dark matter haloes of galaxies with masses similar to or greater than that of the Milky Way are filled with gas that has been shock-heated to temperatures of order or greater than $10^6$ K (e.g. Birnboim & Dekel 2003; Keres et al. 2005; van de Voort et al. 2011; van de Voort & Schaye 2012). This hot circum-galactic medium may contain many more baryons than the galaxies themselves and also harbour a substantial fraction of the heavy elements in the universe (e.g. Tumlinson et al. 2011).

In the absence of feedback processes, much of the hot halo gas would cool and condense onto galaxies, fuelling star formation at a level that is much higher than observed. It is widely believed that outflows driven by feedback from massive stars and/or active galactic nuclei are the solution to this overcooling problem.

Indeed, observations of starburst galaxies consistently reveal the presence of low-ionisation absorption lines, blueshifted by hundreds to thousands of km/s relative to the galaxies' systemic velocities, as would be expected if the galaxies drive galactic winds into intergalactic space (e.g. Veilleux et al. 2005). However, these absorption line measurements can only paint a one-dimensional picture and they do not probe the hot gas that is thought to drive the outflows and to fill the haloes.

**The formation and evolution of galaxies and the evolution of the gas around them are thus closely intertwined.** Galaxies are fuelled by the accretion of gas from their gaseous haloes and they regulate their growth, and that of their central black holes, by driving large-scale winds back into intergalactic space (see Cappi, Done et al. 2013, *Athena+* Support Paper).

The importance of these winds is not limited to models of galaxy formation. Our lack of understanding of the ejection of baryons from the (inner) haloes of massive galaxies may well become the main bottleneck for upcoming precision cosmology experiments based on gravitational lensing or measurements of the halo mass function (e.g. van Daalen et al. 2011, Semboloni et al. 2011). For instance, they show that AGN feedback models predicts a decrease in power relative to a dark matter only simulation ranging, at z=0, from 1% at 3 Mpc scales to 30% at 0.1 Mpc scales. This leads to significant biases in the inferred cosmological parameters such as $w_0$ (~40% bias).

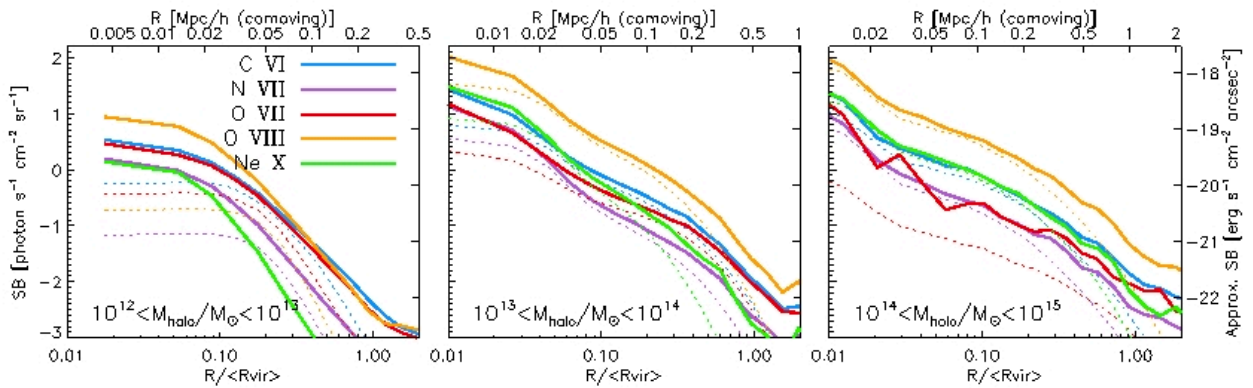

Figure 6: Mean (solid curves) and median (dotted curves) surface brightness as a function of radius at z = 0.125 for the indicated soft X-ray lines and three different halo mass bins. The thickness of the slices is 1.7, 3.7, and 5.6 comoving $h^{-1}$Mpc, respectively, corresponding to four times the largest virial radius in each mass bin. The radius on the bottom x-axis is normalised by dividing by the median virial radius in each mass bin. The surface brightness in photon $s^{-1}cm^{-2}sr^{-1}$ indicated by the left y-axis is exact. For the right y-axis it has been converted to erg $s^{-1}cm^{-2}$arcsec$^{-2}$ using <λ>=22.2 Å (0.558 keV). The pixel size was 5" (8.9 comoving $h^{-1}$kpc; 7.9 proper $h^{-1}$kpc) before binning. O VIII is the brightest line for all halo masses, followed by C VI. The relative strengths of the lines vary with halo mass. The profiles flatten at R≤10 $h^{-1}$kpc, because this region is dominated by the ISM (excluded from this analysis).

For galaxies more massive than the Milky Way, the temperatures that are characteristic for the gas that has passed through an accretion or wind shock correspond to the soft X-ray band. Simulations predict that the X-ray emission from the diffuse halo gas around massive, disk galaxies is close to the detection limit of existing facilities (Crain et al. 2010) and this emission may indeed already have been detected (Anderson & Bregman 2011). With *Athena+* we will move from marginal global detections for the most massive galaxies to spatially resolved images of the circumgalactic media of normal galaxies.

Indeed, many metal lines are expected to be sufficiently strong to be detected in the circumgalactic media of galaxies (e.g. Bertone et al. 2010; van de Voort & Schaye 2013). Because dense, compact structures can easily dominate the emission even if they contain only a very small amount of mass, it is important for future observatories to have high angular resolution. Without sufficient angular resolution, one risks misinterpreting the origin of the detected emission. For example, a large-scale filamentary structure that looks like it is part of the diffuse cosmic web, may in reality come from a number of compact gas clouds in and around galaxies that trace the underlying large-scale structure. Bertone et al. (2010) used hydrodynamical simulations to show that the angular resolution of *Athena+*, ~5", is optimal.

*Athena+* has a very high sensitivity for weak, diffuse line emission. At low energies, the background is dominated by the line emission from the cosmic X-ray background. For square boxes of 5"x5" and 1 Ms exposure time the continuum emission of the background, in the spectral bands free from strong background emission lines, corresponds to only 2.1 and 0.7 counts per resolution element of 2.5 eV, for energies of 0.5 and 1 keV, respectively. Emission lines with a surface brightness of 0.1 photons/s/cm²/sr produce in the same 5"x5" boxes and within the same 1 Ms exposure time 5 and 8 counts, respectively, hence are well above the background.

Fig. 6, taken from van de Voort & Schaye (2013), shows mean surface brightness profiles predicted by the OWLS hydrodynamical simulations for some of the strongest X-ray lines from haloes at redshift z=0.125 with total masses typical of galaxies ($10^{12.5}$ $M_\odot$; left panel), groups of massive galaxies ($10^{13.5}$ $M_\odot$; middle panel), and clusters of galaxies ($10^{14.5}$ $M_\odot$; right panel). Note that these predictions are conservative in the sense that all gas with densities similar to that of the interstellar medium ($n_H > 0.1$ cm$^{-3}$) was excluded.

Fig. 7 illustrates that *Athena+* will e.g. be able to detect O VIII (654 eV) out to at least 80% of the virial radius ($R_{vir}$) of groups and clusters and out to at least $0.4 R_{vir}$ for galaxies. Stacking galaxies or averaging over multiple resolution elements will enable the detection of diffuse halo gas to much greater distances. Other lines, including C VI (367 eV), N VII (500 eV), O VII (561 eV) and Ne X (1021 eV), will also be detectable throughout large parts of the haloes.

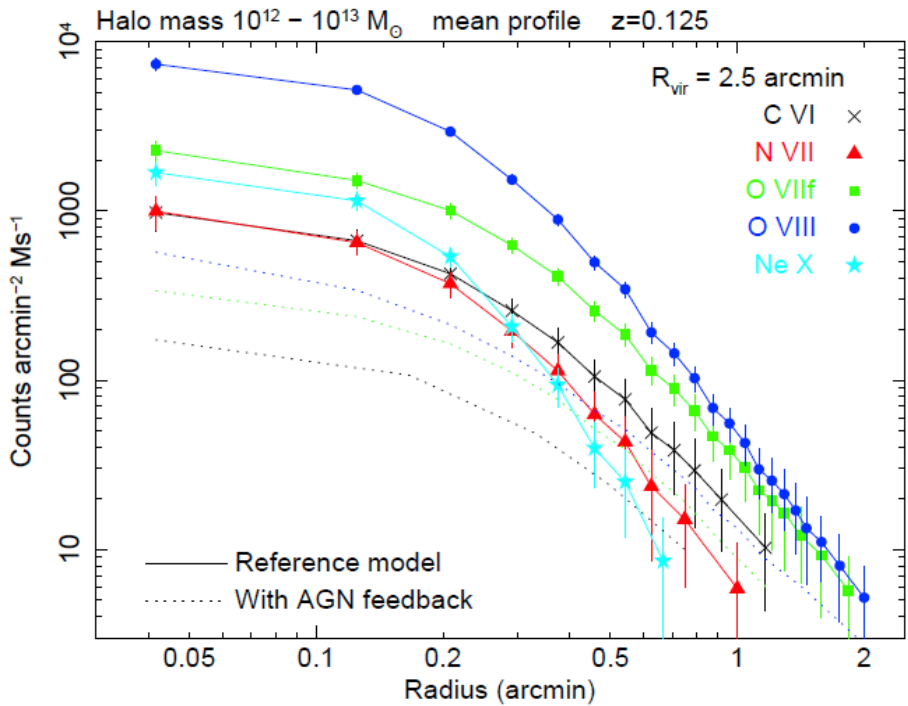

Figure 7: Simulated surface brightness profiles for the 1s-2p transitions of the indicated ions in galactic haloes. Exposure time: 1 Ms, bin size 5". The curves are based on the profiles of Fig. 6 (left panel, reference model) and similar calculations including AGN feedback (Van de Voort & Schaye 2012, Fig. 6). *Athena+* will constrain feedback models in great detail.

The detection of multiple metal lines will allow detailed studies of the physical conditions of the gas, including its temperature, its chemical composition and clumpiness. *Athena+* will provide spatially resolved images that will map the structure of gas flows around individual galaxies and will allow these flows to be related to the rate of star formation and black hole activity and to the orientation of the galaxies.

Thus, *Athena+* will open a unique window onto two of the most important, poorly understood processes regulating the formation and evolution of galaxies: gas accretion and galactic winds.

## 5. FINAL CONSIDERATIONS

The baseline option for *Athena+* is a spectral resolution of 2.5 eV for the XMS detector. Studies are ongoing to see if a part of the array can have a resolution of 1.5 eV (with a modest 20% reduction in effective area). Such an improvement would be important for all the science topics presented here. In particular for the detection of very weak absorption lines, such as those of the WHIM, limited by systematics, the sensitivity will increase by a factor of 170%. For other studies the sensitivity for measuring fluxes, centroids and widths of weak absorption lines would increase by 15, 50 and 250%, respectively.